\documentstyle[12pt]{article}
\title{Phenomenon of electrization caused by gravitation of massive body}
\author{D.A. Kirzhnits, A.A. Shatsky }
\date{}
\begin{document}
\large
\maketitle

  The value of excess charge in the kernel of massive body

  (and the opposite in sign excess charge at the surface)

 caused by the influence of gravitational forces is determined.\\

Up to now an old problem of connecting rotatoty and magnetic
characteristic heavenly bodies has been of gread interest. One of
the aspects of the above problem is a question concerning 
quasi-neutrality disturbance of the substance of the body.
(neutrality which on average in scale is greater than the lattice 
parameter or Debay radius because of gravitational forces.
In addition to condensing the substance of the body these forces
also create an excess of a positive charge in the center of a heavenly
body and an excess of a negative charge on periphery of a body. 
(gravitational forces affect more strongly heavy nuclei than 
 light electrons).
The electric field caused by this effect results in the compression of 
the electric component of the substance of a heavenly body. 
        
      Rotating around its own axis that body owing to redistribution
of its charge must obtain magnetic moment. Thus far the scientists have
tried to explain (by the mechanism under discussion) the magnetic
properties of  heavenly bodies. The aim of  this paper is to prove
hopelessness of such attempts due to a negligible value of the
effect of charge redistribution which corresponds to parameter $\alpha$:
$$\alpha ={Gm^2_p\over e^2}\approx 10^{-36}, \eqno (1) $$ 
Here $G$ - a gravitation constant, $m_p$ and $e$ - the mass and 
charge of  a proton. 

        1.  Let's start with elementary evaluations directly concerning 
a solid (crystalline) state of a substance. In this
case redistribution of a charge is the displacement of a nucleus 
relative to the center of 
"Vigner-Zeitc cell" and leads to the appearance of dipole moment and
polarization [1]. It is obvious that both the electric and gravitational
forces which act on the displacement of a nucleus must be in balance. 
Hence 
$$Ze{\bf E}=-GM^2\nabla \int d{\bf x'} n_p({\bf x'})/{\bf |x-x'|},
\eqno (2) $$
Here $Ze$ and $M$ - a charge and mass of a nucleus, ${\bf E}$ - the
 vector of an electric field, $n_p$ - the concentration of nuclei.
Taking into
account equation $div{\bf E}=4\pi\delta\rho $ ($\delta\rho$ -
an excess of  a charge), one may find: 
$$\delta\rho = ({A\over Z})^2\alpha\rho_p \sim \alpha\rho_p .
\eqno (3)$$
Here $A$ - the mass of a nucleus divided by the mass of a proton, 
$\rho_p$ - the density of an electric charge of nuclei.

Simple dimension considerations are in favour of evaluation $(3)$. 

Dimensionless relation $\delta\rho /\rho_p$ must be proportional to $G$
(or more exactly to $\alpha$) which follows from the application 
of  perturbation theory on the gravitation interaction.
Corresponding coefficient
of proportionality may depend  on  dimensionless  parameters  of  
substance  -  $z$, the ratio of  masses of a proton and an electron;
the ratio of  
Coulomb energy of a single particle to the largest of the following values:
$E_F$ and $T$ ($E_F$ - Ferme energy and $T$ - temperature ). Now you 
can see for yourself that for astronomic bodies the above parameters differ
from 1 not more than 100 - 1000 times.
Therefore they can't considerably influence eq. (3), (because
 $\delta\rho /\rho_p$, depends on $G$ linearly).  
The negative charge, which compensate  (3) and is localized on 
the surface of a body, equals  
$$Q=-\int d{\bf x}\delta\rho\sim\alpha Q_p, \eqno (3')$$
Here $Q_p$ - the total charge of the nuclei of a body. The polarization
${\bf P}=Zen_p{\bf\delta}$ (${\bf\delta}$ - the shift of a nucleus) is equal
to $-{\bf E}/4\pi$, because induction ${\bf D}= 0$ owing  to the 
absence
of external charges. As a result the elementary calculation using eq. (3)
gives the ratio of the magnetic moment of a body to its mechanical
 moment. 
$$\sim -\alpha {e\over m_pc}.\eqno (4)$$
Owing to the infininitesimal of parameter $\alpha $ (see $(1)$) 
the above mentioned values are very small and the effect of
charge redistribution under discussion can't have direct observing 
demonstrations.

	Suffice it to say, that for the Earth (the mass $\sim 10^{27} g.$,
the radius $\sim 10^9 sm$) the magnitude of the surface charge $(3')$ 
corresponds to one electron per 1 $m^2$ of the surface. 

	2. A more rigorous derivation of eq. (2) for a solid state of a
substance is 
based on the selection of the part system from the total energy 
 depending on the shift of nuclei $\delta_k$ (k -  the number of a
 nucleus)
and minimization of this part over $\delta_k$ accompanied by a change 
of: $\delta_k\longrightarrow {\bf p_k}/(Zen_p)$. 
Besides the sums for the grate may be replaced by integrals: 
$$\sum_k\longrightarrow \int d{\bf x}n_p,\makebox{    }{\bf p_k}
\longrightarrow {\bf P(x)}.$$
Let's start with gravitational energy for nuclear interaction :
$$E_{gr}=-{GM^2\over 2}\sum_{k,k'}|{\bf x_k-x_{k'}}|^{-1}$$
Substituting ${\bf x_k\longrightarrow x_k+\delta_k}$ and 
factorizing over $\delta$ up to the first order inclusive, one finds:
$$ E_{gr}=-{GM^2\over Ze}\int d{\bf x}\{{\bf P(x)}\nabla\int 
{d{\bf x'}n_p({\bf x'})\over |{\bf x-x'}|}\}. \eqno (5)$$
Coulomb energy of interaction of nuclei and 
electrons can be written as: 
$$E_c={Ze^2\over 2}\sum_{k,k'}'|{\bf x_k-x_k'}|^{-1}-Ze^2
\int d{\bf x}n({\bf x})\sum_k |{\bf x_k-x}|^{-1}+{e^2\over 2}\int\int
{d{\bf x}d{\bf x'}n({\bf x}) n({\bf x'})\over |{\bf x-x'}|},$$
here $n\approx Zn_p$ - the concentration of electrons. 
In repeating the same considerations which resulted
in (5) we must have in mind
two circumstances:

1.  Now we must factorize ${\bf\delta}$ (or 
${\bf P}$) to the
second degree inclusive, (as $\alpha$ is very small, see  (1)).

2.  Factorization with respect to $\delta$ is impossible
in case of interaction with
electrons of the same cell because of the necessity to take into account the
contribution of region $r<\delta$ and this case  
must be considered exactly.  

It is represented in the second term of $E_c$, in which one
must select an integral over the cell volume:
$$-Ze^2\int {d{\bf x}n({\bf x})\over |{\bf x-\delta}|},$$
multiplied by the total number of nuclei. It leads to the
following expression for the part
 depending on ${\bf P}$:
$$E_c^{(1)}={2\pi\over 3}\int d{\bf x P^2}.$$
In the remaining part of $E_c$ after  factorizing over $\delta$ 
an ordinary term of dipole-dipole interaction
arises, the interaction of different cells being 
reduced to it:
$$E_c^{(2)}=\int\int d{\bf x}d{\bf x'}({\bf {(P(x)P(x'))\over 2|x-x'|^3}
-3{(P(x),P(x-x'))(P(x'),(x-x'))\over 2|x-x'|^5}})$$
(Terms of $E_c$ which are linear with respect to
 ${\bf P}$ disappear  owing to the
neutrality of a system).  The equation for $E_c^{(2)}$ may be written as 
($see$  $ Enclosure $):
$$E_c^{(2)}={2\pi\over 3}\int d{\bf x [P^2(x)-3(P(x),\nabla){(\nabla
 P(x))\over \Delta }}]$$
For that reason the sum of $E_c^{(1)}$ and $E_c^{(2)}$ can be written as:
$$E_c=2\pi\int d{\bf x P^2_l}, \eqno (6)$$
here ${\bf P_l}={\nabla div\over \Delta}{\bf P}$ - the longitudinal 
part of vector ${\bf P}$ (it is in fact in $(5)$).
Minimization of sums $(5)$ and $(6)$ over ${\bf P_l}$
(if ${\bf E}=-4\pi{\bf P}$) brings us back
to eq. (2).  

Let's emphasize  that in the last equation there 
is no correction for
the distinction of the functioning field from the 
average one [1].  That
correction would appear if we placed the nucleus
considered - for which the balance of forces
is written - into empty space (2). In fact 
we have the interaction of a nucleus with electrons of its cell (cavity). 
It is described by $E_c$ and the required
 contribution $-4\pi{\bf P}/3$ to intensity
because ${\delta E\over \delta{\bf P}}=-{\bf E}$. 
To complete the proof  of eq. (2) one should make sure that 
the parts of the system energy not considered above 
(to be more exact their parts depending 
on ${\bf P}$) do not influence the result. First of all it is true  for the
electronic component of energy - kinetic, exchange, correlation [2].  

	In case of a strongly compressed substance - this case is more
interesting for heavenly bodies having a solid state substance -  
kinetic energy of free electron gas is an important factor (other
 components of energy are very small compared with Coulomb energy
taken into account above).  

     Decomposition of this energy over the nuclear shift relative to the
center of a cell:
$$\delta E_{kin}\sim \int d{\bf x}n\delta U, \makebox{      } 
\delta U= Ze^2({1\over r} - {1\over |{\bf r+\delta}|})$$
leads to the zero result owing to the spherical symmetry $n$ and 
a known decomposition of Coulomb term into a series over 
Lezhandr polynomial. As regards the energy of nuclei in a solid 
state at low temperatures one should consider only zero  energy
of nuclear oscillation which is equal to 
${3\over 2}\hbar w_0 $ for one nucleus ($w_0$ - 
the oscillation frequency). Hence only part $w_0$ depending 
considerably on shift  $\delta $ or ${\bf  P}$ could affect the proof given
above. But in case of an exceptional small value of $\delta $ 
($\delta \sim \alpha R$ - see point 1, $R$ - the radius of a body)  the shift 
of  the oscillation center of a nucleus in the cells model does not 
influence the oscillation frequency at all. A change in energy in shifting a
nucleus by ${\bf\delta_r}$ is equal to $l^2e^2\delta^2_r/(2R^3)$ 
(where $R$ - the radius of a cell) and in substituting 
${\bf  \delta_r \longrightarrow\delta_r + \delta}$  (the shift
of the oscillation center) the square of frequency as the second 
derivative of energy with respect to $\delta_r$ does not change at all.

   3. Eq. (3) is in fact of universal nature and it is valid irrespective
of the state of a substance. We will utilize a method 
of functional density [3], writing down the free energy of a system 
(in general case temperature $T$ is not equal to zero) as a function 
of the density of electrons $n$ and nuclei $n_p$.
$$F\{n,n_p\}=F_0+E_{sc}+F_{xc}-\mu \int d{\bf x}n -
 \mu_p \int d{\bf x} n_p , \eqno (7)$$
here the first term conforms to free electronic and nuclear 
gases, the second one - Coulomb and gravitation energy of their
interaction in approximation of a self - consistent field, the third term 
corresponds to exchange and correlation effects. The latter two terms
correspond to Lagranzh factors and show conservation of  
total numbers of electrons and nuclei. Minimum $F$ for $n$ 
and $n_p$ defines equilibrium distributions of these quantities. Writing
down 
$$E_{sc}=-{GM^2\over 2}\int {d{\bf x}d{\bf x'}\over {\bf |x-x'|}}n_p 
n_p ' +{e^2\over 2}\int {d{\bf x}d{\bf x'}\over {\bf |x-x'|}}(n-zn_p)(n'-
zn_p '), $$
we obtain conditions of minimum $(7)$ for $n$ and $n_p$:
$$\Delta {\delta (F_0+F_{xc})\over \delta n({\bf x})}=4\pi e^2 (n-zn_p) 
\eqno (7')$$
$$\Delta {\delta (F_0+F_{xc})\over \delta n_p({\bf x})}=
-4\pi z e^2 (n-zn_p)-4\pi GM^2 n_p. \eqno (7'')$$

If we could omit the left part of eq. $(7'')$, then, taking into 
account that $\delta \rho =e(zn_p-n),\makebox{ }\rho_p =zen_p $, we 
would obtain eq. $(3)$, and if $(7'')$ is substituted into $(7')$ and if
appropriate simplifications are made then $(7')$ will fit Chandrasekar 
equations which define equilibrium configuration of electrons
and nuclei. Let's emphasize that the left part of that equation is defined 
by the lightest particles - electrons, the right part - contains only
 gravitation quantities after the 
above substitution of $(7'')$ into $(7')$
although gravitation does not act on the electrons directly. The electrons
are affected by the electric field investigated in this paper, which  
only quantitatively (see $(3)$) coincides with a gravitational field. 

Thus eq. $(7')$ and $(7'')$ are rewritten as:
$$\delta \rho =\alpha\rho_p (1+\sigma )^{-1};\makebox{  }\sigma 
={\Delta\delta F/\delta n_p\over z\Delta\delta F/\delta n},\eqno (8)$$
where here and below $F=F_0+F_{xc}$.

The appearance of $\sigma $ either keeps the magnitude of 
$\delta\rho /\rho_p $ constant, or decreases this relation. The single case
when it may increase considerably, represents exceptional 
nearness of $\sigma $ and $-1$. But this is practically unbelievable.
Furthermore, we may think that $\sigma <<1$ . Let's illustrate it by
using two examples [2]. For both examples it is supposed that the 
electron gas is strongly compressed and degenerated so that the
 corresponding contribution to $F$ is $\sim {\hbar^2\over m}\int 
d{\bf x}n^{5/3}$. In the first example nuclei are localized in nodes 
of a grate and the energy of their zero oscillations represent 
$\delta F\sim {Ze\hbar\over \sqrt{M}}\int d{\bf x}n^{3/2}_p$ (see 
above). Then for $\sigma$ in $(8)$ we find: 
$$\sqrt{{m\over M}Z}(a_0n^{1/3})^{-1/2}<<1,$$
where $m$ - the mass of the electron, $a_0=\hbar^2/me^2$ - it's 
"Bohr-radius". This smallness is connected with inequalites: 
$m/M<<1$ and $a_0n^{1/3}>>1$ in a compressed substance. 

The second example represents a weak non ideal nuclear 
"Boltzman-system"  for which $F(n_p)\sim -e^3\int
d{\bf x}{n_p^{3/2}\over \sqrt{T}}$ is "Debay-Hukkel" - correlation 
correction. So for $\sigma $ one finds :
${e\sqrt{n^{1/3}/T}\over (\sqrt{Z}a_0n^{1/3})}<<1$
because the condition of weak non-perfection is $T>>e^2n^{1/3}$.

4. In conclusion let's go back to the question about the minimum 
of  $(7)$ in connection with disturbance of local electric neutrality of the 
system. 

 Note that the above violation is typical of the crystalline state of a
substance in the absence of  gravitation forces, which is evident
and the electrons are not localized in contrast to nuclei which
are localized at the point. 

It is important that this violation is not described by minimum of
F, as a point in functional space in which its functional derivative 
disappears. In this case we deal with a boundary minimum
reached when the parameter which defines the length of nuclear 
localization tends to its limiting value which equals zero. 

Let's consider the second clear Coulomb component of 
$E_{sc}$ which should be broken into neutral and spherical on the 
whole "Vigner-Zeitc" cells with a 
nucleus in the center. In such a model the radius of the cell is $R$
and the nucleus is spread in sphere with radius $\rho$. This model
represents energy: $E_{sc}=-{3\over 10}{Z^2e^2\over R}{2x^3+4x^2+6x+3\over 
(x^2+x+1)^2}$
where $x=\rho /R$. At  $x=0$ this equation gives a well-known 
binding energy of the grate: $-{9Z^2e^2\over 10R}$. 
It is evidentl that maximum of  $|E_{sc}|$ is in fact reached on the 
boundary of permissible area (at $x=0$ and fixed $R$).
$$ $$
$$\makebox{                                          } Enclosure $$
Initial expression for $E_c^{(2)}$ can be written as:
$$E_c^{(2)}=\int d{\bf x}P_i({\bf x})K_{ij}({\bf k})P_j({\bf x}),$$
where ${\bf k}=-i\nabla ,\makebox{  }\nabla $ act on $P_j({\bf x})$,
and
$$K_{ij}(\vec k)={1\over 2}\int {d\vec\lambda\over\lambda^3}(\delta_{ij}-
3{\lambda_i\lambda_j\over\lambda^2})e^{-i\vec k\vec\lambda}.$$
This integral satisfies evident condition $K_{ii}=0$ and hence 
it may be represented as 
$$K_{ij}=-{1\over 2K^2}k_lK_{lm}k_m(\delta_{ij}-{3k_ik_j\over
 k^2}).$$
The expression for $E_c^{(2)}$ mentioned in this paper gives the 
calculation that is not complicated but uwkward.
$$ $$

In conclusion we would like to note critical debate on this questions 
with Vasilyev B.V., Grigoryev V.I. and Maximov V.I..
$$ $$

Lebedev Institute of Physics RAS,

Physics Department, Moscow State University.
$$ $$
$$\makebox{                                 } LITERATURE$$
$$ $$

1. Tamm  I.E. The Theoretical Course on Electricity. Moscow, Science, 
1989.

2.  Landau L.D., Lifshitc E.M.. Statistical physics, part 1, Moscow, 
Science, 1995. 

3.  Lundkvist S., March M., The Theory of Heterogeneous Electrons gas. 
Moscow, Mir, 1987. 

\end{document}